\documentclass[aps,onecolumn,10pt,final, showpacs]{article}
\usepackage{amsmath,amssymb,amsfonts}
\usepackage{graphics,graphicx}
\usepackage[unicode,colorlinks,linkcolor=blue]{hyperref}
\usepackage{verbatim}
\usepackage{latexsym}
\usepackage{subfigure}

\title{Inflationary limits on the size of compact extra space}
\author{V.V.~Nikulin \\ National Research Nuclear University MEPhI \\ (Moscow Engineering Physics Institute),\\ 115409, Kashirskoe shosse 31, Moscow, Russia \\
Sergey G. Rubin \\
    National Research Nuclear University "MEPhI", \\ (Moscow Engineering Physics Institute),  \\
    N.I. Lobachevsky Institute of Mathematics and Mechanics,\\
    Kazan  Federal  University, \\
    Kremlevskaya  street  18,  420008  Kazan,  Russia\\
    sergeirubin@list.ru}

\begin{document}
    
    \maketitle
    
    \begin{abstract}
We study restrictions imposed on the parameters of the Kaluza-Klein extra space supplied by the standard inflationary models. It is shown that the size of the extra space cannot be larger than $\sim 10^{-27}$cm and the $D$-dimensional Planck mass should be larger than $\sim 10^{13}$GeV. The validity of these estimates is discussed.

We also study the creation of stable excitations of scalar field as the result of the extra metric evolution.
    \end{abstract}
    
        \section{Introduction}
The inflationary period in the early Universe \cite{Starobinsky:1980te,Lindebook} is the widely admitted paradigm that has passed all observational tests. The inflationary period is successful in explaining the spatial flatness, homogeneity, and isotropy of the Universe as well as the temperature fluctuations of the Cosmic Microwave Background radiation and large-scale cosmic structures. In the following, we will rely on this paradigm as a firmly established fact.
        
There are two main processes during inflation - quick space expansion and growth of the field energy due to fluctuations. In simplest and most widespread realization of the inflationary scenario, only one-dimensional parameter plays an essential role - the Hubble parameter $H$. In the following, we assume the value of the Hubble parameter $H\simeq 10^{13}$GeV which relates to the scale $l_h\sim 10^{-27}$cm, the size of the horizon at the inflationary stage.

If a compact space is involved into consideration, another dimensional parameter - an extra space size $l_d$ - appears. There is a lot of models describing the extra dimensions with various metrics and topologies, for example \cite{2011PhRvD..84d4015B,1999PhRvL..83.3370R,hall2001gauge,rubakov2001large}. The extra space that which affects the parameters of our Universe can have a complex structure consisting of subspaces with different properties. Some of them could be well known compact subspaces of the Kauza-Klein (KK) type. In this paper, we study this type of subspaces.

d-dimensional compact extra space is characterized by two main parameters - the D-dim Planck mass $m_D$ and a scale $l_d$ of compact extra dimensions. Both of them varies in range from the Planck scale to the scale $\sim 1 $TeV for $m_D$ and to $\sim 10^{-18}$cm for  $l_d$ depending on a specific model. The hope to find them at the LHC collider is based on these numbers, see \cite{Deutschmann2017CurrentDimensions,2018arXiv180808801G}. At the same time, there exists another collider created by Nature itself - the early Universe, or more definitely its first stage of evolution - the inflation. 
    
There are two opposite cases  $l_d\gg H^{-1}$ and $l_d\lesssim H^{-1}$. The first case leads to the energy storage "inside"\ the extra space in the form of its fluctuations. It will be shown here that this contradicts the observations. Therefore,  if the compact space does exist its size should be smaller than  $\simeq H_I^{-1}\simeq 10^{-27}$ cm. This limit is much more strong as compared to those obtained from the  LHC collider. If the second case $l_d\lesssim H^{-1}$ holds, the KK metric excitations are too massive be excited during the inflation.  

In Sections 2 and 3 we assume that a ground state of the extra space is described by a stationary metric. The metric variation is discussed in Section 4.

\section{Inflation and limit on the size of extra space. }

In this section, we discuss the influence of compact extra dimensions on the properties of the inflationary stage. A particular model of inflation does not matter. The facts which only will be used is that the 4-metric is almost the de Sitter one and the Hubble parameter is almost a constant during the inflation. These conditions are met for a wide set of models \cite{martin2014encyclopaedia,akrami2018planck}. It will be shown that the size of the extra space must be smaller than the inverse Hubble scale during inflation for not to contradict the inflationary features mentioned above.

Consider a $D$ dimensional action
\begin{equation}\label{act2}
S=\frac{m_D^{D-2}}{2}\int d^4 x d^d y \sqrt{|g|}\left[R_D+\frac12 \partial_A\chi g^{AB}  \partial_B \chi - \frac12 \mu^2\chi^2 -2\Lambda_D\right]
\end{equation}
for a scalar field $\chi(x,y)$ which plays the role of the inflaton.  $m_D$ is the $D$-dimensional Planck mass.
The metric is chosen in the form
\begin{equation}
ds^2=dt^2-a(t)^3 dl_3^2-l_d^2d\Omega_d^2,\quad l_d=const,
\end{equation}
where $l_d$ is a radius of the maximally symmetric $d$-dim extra space. The choice of such kind of metrics strongly facilitates the analysis and widely used in the literature.

To proceed, let us reduce \eqref{act2} to the 4-dim effective action. The standard way is to expand the function $\chi(x,y)$ in term of the eigenfunctions $Y_q(y)$
\begin{equation}\label{sum}
\chi(t,x,y)=\sum_{q}\chi_q (x)Y_q(y).
\end{equation}
where
\begin{equation}\label{Yq}
\square_d Y_q(y)=M_q^2 Y_q(y),\quad \frac{1}{v_d}\int d^d x \sqrt{|g_d|} Y_q(y)Y^*_p(y) =\delta_{qp}.
\end{equation}
Integration out of the extra coordinates $y$ leads to the action in the form
 \begin{equation}\label{S5}
S=\int d^4 x\sqrt{-g_4}\left[\frac{M_P^2}{2}R_4+\sum_q\left(|\partial_{\mu}\chi_q|^2-(M_q^2+\mu^2)|\chi_q|^2\right)\right]
\end{equation}
Here denotations are
\begin{equation}\label{denot0}
M_P^2= v_d m_D^{d+2};\quad \Lambda_D =\frac12 m_D^2d(d-1)l^{-2}_d;\quad M_q=\frac{q}{l_d}; \quad q=0,1,2...
\end{equation}
where $q$ enumerates the excited states.
We will neglect in the following the bare mass $\mu$ as compared to the Kaluza-Klein mass tower $M_q$ for simplicity. The more general case leads to the same conclusion. The effective potential looks as follows
\begin{equation}\label{pot0}
V=\sum_{q=0}^{\infty}\frac12 M^2_q\chi_q^2(t).
\end{equation}
There are a great variety of models describing the first stage of our Universe formation \cite{martin2014encyclopaedia,akrami2018planck}. One of the most known branches of them is based on the chaotic inflation with the slowly moving scalar field(s). The latter is the reason for the rapid, almost exponential space expansion while the horizon size remains unchanged if measured in the physical coordinates. 
This means that the space is divided into quickly growing number of disconnected areas. Each space volume is limited by its own horizon $l_h=H^{-1}$.  The value $l_h\simeq 10^{-27}$cm and consequently $H\simeq 10^{13}$GeV $\simeq$ const are in agreement with observations \cite{akrami2018planck}. 

It will be shown in this Section that the process of inflation described above cannot be realized if the size of compact extra space $l_d$ is larger than the horizon size. More definitely, we show that if
\begin{equation}\label{H}
    H\simeq const
\end{equation} and
\begin{equation}\label{ineq1}
n=\frac{l_d}{l_h} =l_d H \gg 1
\end{equation}
we come to a contradiction.

According to the chaotic inflation \cite{Lindebook} for the action \eqref{S5}, the fields evolve as
\begin{equation}\label{sol}
\chi_q (t)=\chi_{q,in}e^{-\frac{M_q^2}{3H}t}
\end{equation}
The KK tower contains light modes ($M_q <H$) and heavy modes ($M_q >H$). The latter goes to the potential minimum before the inflation is started. The light modes of the scalar field are active during inflation. The number of modes is limited from below according to the inequality 
\begin{equation}
M_q <H\rightarrow q<l_dH\equiv n
\end{equation}
The upper limit on the extra space size that has been reached at LHC is $l_d\sim 10^{-18}$cm. In this case, the number of active modes is $n\sim l_dH\sim 10^{9}$.  Therefore, sum \eqref{pot0} contains $10^9$ terms and may be approximated by the integral. For further estimation, suppose that the initial field values do not depend on $q$:
\begin{equation}\label{qind}
    \chi_{q,in}\simeq \chi_{in}
\end{equation}
It gives the following estimation of the potential in \eqref{pot0} by the integral
\begin{eqnarray}
&& V=\sum_{q=0}^{\infty}\frac12 M^2_q\chi_q^2(t)\simeq M^2_1\chi^2_{in}\int_{0}^{\infty}e^{-\frac{2M_1^2q^2}{3H}t}q^2dq \nonumber \\
&& =\frac14 \chi^2_{in}M^2_1 \left(\frac{3}{2M^2_1}\right)^{3/2}\left(\frac{H}{t}\right)^{3/2}, \nonumber
\end{eqnarray}
and we arrive to the equation for the Hubble parameter
\begin{equation}\label{Hubble2}
H=\sqrt{\frac{8\pi}{3M_{Pl}^2}V}\simeq  \sqrt{\frac{\pi}{M_{Pl}^2}\chi^2_{in}\left(\frac{3}{2M^2_1}\right)^{1/2}\left(\frac{H}{t}\right)^{3/2}}
\end{equation}

Solving this algebraic equation for $H$ we obtain
\begin{equation}
H\sim t^{-3}.
\end{equation}
Such time dependence of the Hubble parameter is unacceptable for the inflationary period. Therefore, if inequality \eqref{ineq1} is true the slow roll inflation is absent. We come to the conclusion that the extra space size is bounded above,
\begin{equation}\label{ineq4}
    l_d \leq 1/H \simeq 10^{-27} \text{cm}.
\end{equation}
This is a strong limit for the extra space size. Nevertheless, this is not a firm no-go theorem. There are some ways to escape the contradiction. One of them is to change equality \eqref{qind} which is natural but quite an uncertain estimation. For example, we may choose initial conditions as $\chi_{1,in}\neq 0,\quad\chi_{q,in}=0$ which are realized with small probability. In this case, the standard inflationary scenario may occur.
There are also inflationary models that need not slow rolling \cite{damour1998inflation,linde2001fast}, but they are not very common.

The inequality \eqref{ineq4} allows to impose restrictions also on the $D-$dimensional Planck mass. Indeed, according to \eqref{denot0} 
\begin{equation}
 M_P^2= v_d m_D^{d+2}\simeq l_d^d m_D^{d+2}< H^{-d}m_D^{d+2}
\end{equation}
This leads to the limit in the form
\begin{equation}\label{mDineq0}
    m_D > \left(\frac{H}{M_P}\right)^{\frac{d}{d+2}}M_P.
\end{equation}

    \section{Inflation and limit on the \\
    $D-$dimensional Planck mass}

 In this Section, we argue for a firm limit on the $D-$dimensional Planck mass imposed by the inflationary period of the early Universe. The arguments are quite general so that a specific model of the inflation and a structure of the extra space do not matter. In particular, the extra space metric is not obligatory the maximally symmetrical one. The only what we need is the stationarity of the extra dimensions.
 
Consider the action
\begin{equation}\label{act0}
    S_D=\frac{m_D^{D-2}}{2}\int d^D x \big(R_D-2\Lambda_D+...\big)
    \end{equation}
in $D$- dimensions, $D=4+d$. There are a variety of ways to make the extra dimensions stable like complicated potential of scalar fields \cite{1980PhLB...97..233F,2002PhRvD..66b4036C,2002PhRvD..66d5029N}, the gravity with higher derivatives \cite{2003PhRvD..68d4010G,2006PhRvD..73l4019B,2006A&AT...25..447S,2018GrCo...24..154B} and so on. In this paper we do not discuss ways of the extra space stabilization.   
    
At high energies, the metric $g_D$ of $D$-dim space $V_D$ is quite uncertain. On the other hand, modern physical laws are assumed to be described by a specific stationary metric. As the example, it could be the direct product $V_4\times V_d$ of 4-dim FRW space $V_4$ and a $d$-dim compact space $V_{d}$. Therefore, the transition $V_D\rightarrow V_4\times V_d$ should take place. This happens at the $D$-dimensional Planck scale $l_q\equiv 1/m_D$ where the quantum corrections still dominate. 
There is another parameter that plays a significant role - the Hubble parameter, $H$. The latter decreases with time from a value $H_I$ at the beginning of the inflation to the almost zero value nowadays.   
   
   The size $l_h=1/H$ of a causally connected area at the inflationary stage is the classical value by definition. The classical description of the inflating Universe gives promising results. On the other hand, if the quantum scale $l_q$ is larger than the classical value $l_h$, the latter has no physical sense. This means that inequality
   \begin{equation}\label{mDlim}
       l_q<l_h \rightarrow m_D>H
   \end{equation}
   should take place.
   Knowledge of the Hubble parameter gives the lower limit for the $D$-dim Planck mass,
    \begin{equation}\label{mD13}
    m_D> 10^{13}\text{GeV}
    \end{equation}
The conclusion is that any calculations of $H$ under condition  $l_q >l_h$ (or $m_D <  H$) cause serious question. 

Nevertheless, let us suppose that $m_D\simeq H$ holds somewhere in the middle of the inflationary stage, at the e-fold number $N$ and we formally calculate the horizon scale $l_h$. It gives another interesting possibility. At this moment, we have $e^{3N}$ independent causally connected area where the transition from $V_D\rightarrow V_4\times V_d$ has happened. As was discussed in \cite{2016PhLB..759..622R}, the choice of those $d$ dimensions which should be compactified is random. Therefore, each causally connected 4-dim volume is endowed by their own coordinate directions that have been compactified. It leads to funnels production \cite{2016PhLB..759..622R} that are observed as point-like objects with the mass of the order of the Planck mass. Let us estimate their number. It is known that  $e^{3N}\quad (N\simeq 60)$  causally disconnected areas are formed to the end of the inflation. This means that the total mass of the funnels in our Universe is huge that contradicts the observations and confirms inequality \eqref{mDlim}.

One can conclude that the $D$-dimensional Planck mass $m_D$ is strongly limited from below. There are two lower limits - \eqref{mDineq0} and \eqref{mDlim}.

\section{Violation of the internal charge conservation during inflation}

 It is well known that any symmetries lead to a charge conservation and symmetries of extra spaces are no exception, see for example \cite{Blagojevic,cianfrani2005gauge,Kirillov:2012gy}. The smallness of such charge density in the modern Universe is not very strange. Indeed, suppose that the charge has been formed before the beginning of the inflationary period when the present value of the horizon was very small, $l_{hor}\simeq H^{-1}\sim 10^{-27}$cm.  If the charge is conserved, it is quickly diluted in the course of inflation. 

In this Section, we consider the Kaluza-Klein extra space which has no symmetries during inflation and the associated charge is not conserved. Hence, this charge can be generated at that period. In the post-inflationary period, the final relaxation of the extra space metric leads to its symmetrization and, consequently, to the conservation of the accumulated internal charge.

If several energy levels of the Kaluza-Klein tower can be excited during inflation, a step-wise character of observational spectra must be observed. This could contradict observations. At the same time, if all energy levels except the first one are too heavy, this first level could play an essential role in the inflationary period. Due to the momentum conservation, the lowest KK-excitation is survived if the scalar field is complex (see below). Following this logic, suppose that extra space metric has the form:
\begin{equation}\label{metric}
ds^2 = dt^2 - a(t)^2 (dr^2 + r^2 d\Omega_4^2) -r_0^2e^{2\beta(t,\theta,\phi)}d\Omega_2^2,
\end{equation}
where $\beta(t,\phi,\theta)$ define the geometry of extra space. By assumption, such extra space is symmetrized, $\beta(t\rightarrow \infty ,\theta,\phi) \rightarrow \beta_0(\theta)$, (and as a result, it gains a Killing vector) after inflation.  Independence on the polar angle $\phi$ indicates the presence of isometry along this coordinate
(existence of Killing vector). It is also assumed that time variation of the extra space metric $\beta$ is small during the inflation.

Let us consider 6-dimensional action \eqref{act2} with a scalar field. The charge which is associated with the Killing vector of the extra metric $\beta_0(\theta)$ is expressed in terms of the field $\chi$ as follows:
\begin{equation}\label{Q}
Q(t)=\frac{m_6^{4}}{2}\int d^3x dy_1 dy_2\sqrt{-g} j^0,
\end{equation}
\begin{equation}\label{j0}
j^0=\frac{\partial L}{\partial (\partial_0\chi)} \partial_\phi\chi+ c.c. = \partial^0 \chi^* \cdot \partial_{\phi}\chi+ c.c.
\end{equation}
Here $y_1\equiv \theta,\quad y_2\equiv \phi$. The physical meaning of the charge \eqref{Q} is the internal angular momentum along $\phi$ coordinate.

Let us come back to reduced 4-dimensional action \eqref{S5} that can describe the chaotic inflation. The scale factor has the standard form during the slow roll regime
\begin{equation}\label{slowroll}
a(t)=H^{-1}e^{Ht},
\end{equation}
where $H$ is defined in the first equality of expression \eqref{Hubble2}. Here we assume that only the first KK-level is excited. The opposite case contradicts observations, as was discussed in Section 2. 

Substitution of expressions \eqref{j0}, \eqref{metric}, \eqref{slowroll} into \eqref{Q} gives:
\begin{equation}\label{Delta0}
Q(t)=\frac{m_6^4 r_0^{2} H^{-3}}{2} e^{+3Ht}\int d^3x d\theta d\phi \sin\theta e^{2\beta(t,\theta,\phi)}\left[\partial^0\chi\partial_{\phi}\chi^* + c.c.\right].
\end{equation}
Internal momentum $Q(t)$ varies until the extra metric is distorted (and the symmetry is broken). To the end of the inflationary period a charge $\Delta Q$ is stored. It is assumed that the extra space acquires the $U(1)$ symmetry to the end of inflation and hence the charge is not varied  further. Our goal is to estimate the charge $\Delta Q $, accumulated by the end of inflation. To perform this plan we have to solve the equations of motion:
\begin{eqnarray}\label{eq1}
\square_6 \chi + \mu^2\chi = 0,\quad \square_6 \chi^* + \mu^2\chi^* = 0=0.
\end{eqnarray}

We suppose that $\beta(t,y)\approx \beta(y)$ is almost independent of time during inflation. Following the standard procedure, we decompose the field $\chi$ in the following manner (the coordinates and 3-momentum $k$ are dimensionless)
\begin{eqnarray}\label{decomp}
\chi(t,x,y)=\sum_{q}\chi_q(t,x) Y_q(y), \\
\chi_q(t,x)=\int \frac{d^3k}{(2\pi)^3}e^{i\Vec{k}\Vec{x}} f_q(k,t), \nonumber
\end{eqnarray}
where the eigenfunctions are defined in \eqref{Yq}. The masses of the KK-states are denoted as $M_q=\lambda_q/r_0,\ q=0,1,2...$ . The extra space is not maximally symmetric and the eigenvalues $\lambda_q$ can be found numerically if the metric function $\beta(y)$ is known. Equation \eqref{eq1} acquires the form
\begin{eqnarray}\label{eq3}
&& \partial^2_t f_q + 3H\partial_tf_q + \left(\frac{H^2k^2}{e^{2Ht}}+\mu^2+M^2_q\right) f_q=0.
\end{eqnarray}after substitution the series \eqref{decomp}.  
Equation \eqref{eq3} is known to be complicated \cite{riotto2002inflation}. We will use the solution of the massless equation \cite{Lindebook}
\begin{equation}\label{sol0}
\tilde{f}(k,t)=\frac{1}{\sqrt{2k^3}}(i+ke^{-Ht})e^{ike^{-Ht}}
\end{equation}
to simplify the analysis.
Let us represent the solution as $f_q = C_q(k,t) \tilde{f}(k,t)$ where the function $C_q(k,t)$ is a solution to equation
\begin{equation}\label{eq4}
\partial^2_t C_q +2\partial_tC_q\frac{\partial_t \tilde{f}}{\tilde{f}} +3H\partial_t C_q + (\mu^2 + M^2_q) C_q  = 0.
\end{equation}
Second term is proportional to $\partial_t \tilde{f}/\tilde{f}\sim e^{-Ht}$ and may be neglected after a couple of e-folds. 
The friction is huge during the  inflation so that $C_q$ varies slowly with time and we may neglect $\partial^2_t$ in the spirit of slow motion. As the result, \eqref{eq4} is reduced to the equation
\begin{equation}\label{eq5}
3H\partial_t C_q + (\mu^2 + M^2_q) C_q \simeq 0. 
\end{equation}

The solution to \eqref{eq5} is well known that leads to the solution of equation \eqref{eq3} in the form
\begin{equation}\label{sol1}
f_q(k,t) = A_q \tilde{f}(k,t) \exp \left\{-\frac{\mu^2 + M^2_q}{3H} t \right\},
\end{equation}
where $A_q$ is the initial amplitude of each mode $Y_q(y)$.

\begin{figure}[h!]
    \centering
    \includegraphics[width=0.57\linewidth]{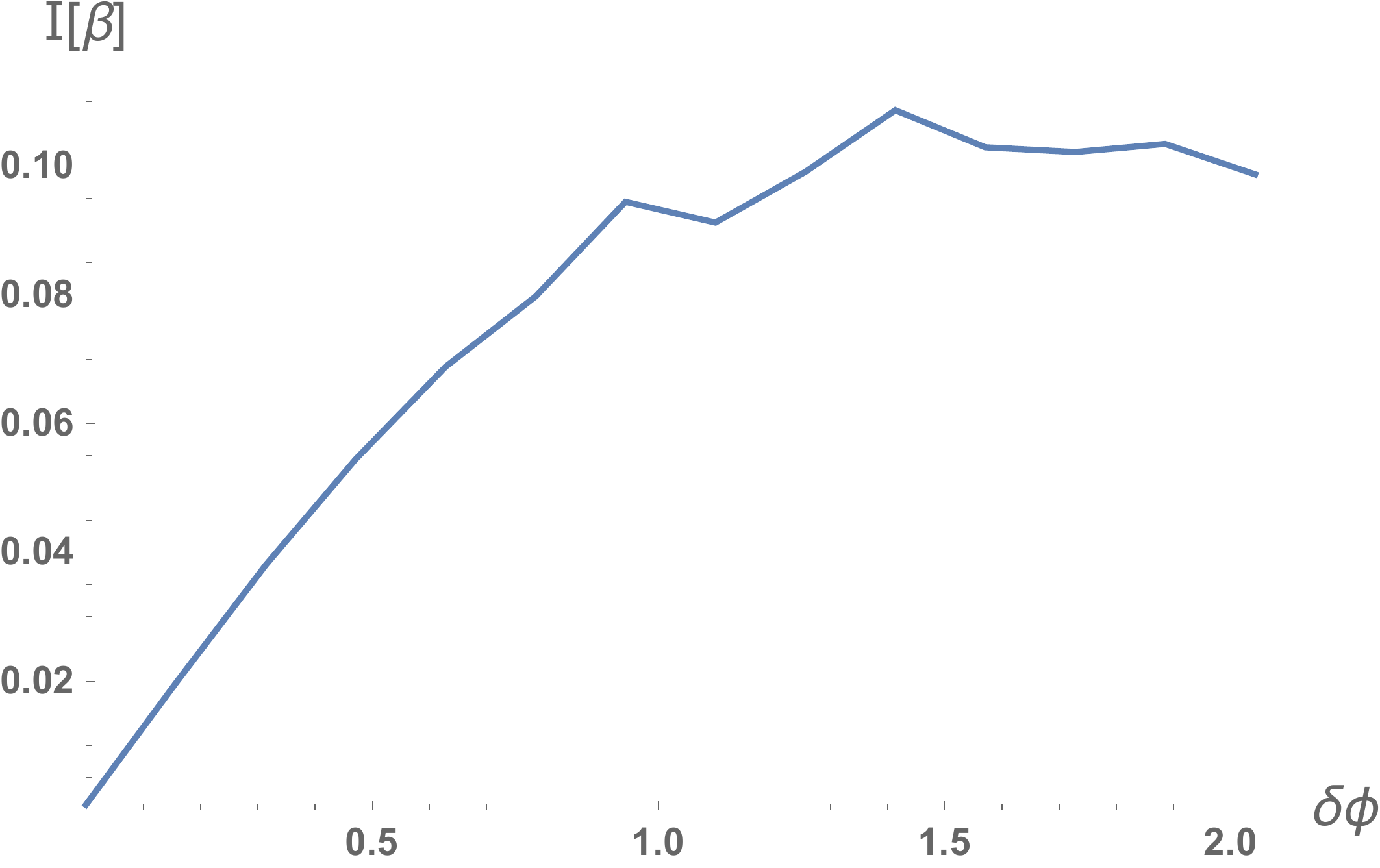}
    \caption{The value of the integral $I[\beta]$. There is a big outrage in the choice of $\beta(\theta,\phi)$. Here we have chosen one of the many possible functions $\beta(\theta,\phi)=\left(0.4\sin(\phi)+0.7\sin(3\phi+\delta\phi)\right) \sin^2(\theta)$ just for the integral estimation.}
    \label{fig}
\end{figure}

Let us estimate the charge. As mentioned above, we consider only the first excitation level. Substituting solutions \eqref{sol0},\eqref{sol1} into \eqref{Delta0} and discarding small terms, we get:
\begin{eqnarray}\label{Delta1}
&&Q(t)=\int\frac{d^3 k}{(2\pi)^3} Q_k(t), \\
&&Q_k(t) = \frac{\lambda_1^2 m_6^4 H^{-4} |A_1|^2}{12 k^3} \exp\left\{3Ht-\frac{2\lambda_1^2t}{3Hr_0^2}\right\} I[\beta]+O(e^{Ht})\nonumber \\
&&I[\beta]=\int d\theta d\phi \sin\theta e^{2\beta(\theta,\phi)}\left[Y_1 \partial_{\phi}Y_1^* + c.c.\right] . \nonumber
\end{eqnarray}
The dimensionless integral $I[\beta]$ can be estimated as $I[\beta] \lesssim 1/\pi$ ($\partial_\phi Y_1 \sim 1/\pi$ because the mode $Y_1$ has the longest wavelength $\sim 2\pi$, and the remaining part of the integral $I[\beta]$ is normalized by condition \eqref{Yq}). Numerical simulation of the functional $I[\beta]$ is represented in Fig.\ref{fig}. The specific form of the function $\beta(\theta,\phi)$ depends on the metric fluctuations before the inflationary stage and is quite arbitrary.

Limits of integration into \eqref{Delta1} are standard for inflationary stage (from $He ^{-Ht}$ to $H$) \cite{Lindebook}. So we obtain the total charge (total internal angular momentum) stored by the field $\chi$:

\begin{equation}\label{Delta2}
Q(t) = \frac{\lambda_1^2 m_6^4 |A|^2 I[\beta] N_t}{24\pi^2 H^4} \exp\left\{3N_t-\frac{2\lambda_1^2N_t}{3H^2r_0^2}\right\},
\end{equation}
where $N_t=Ht$ is the total number of e-folds of the inflationary stage.

The internal angular momentum of a single first-level Kaluza-Klein excitation  $\approx\lambda_1$. It's mass $M_1=\lambda_1/r_0$. Therefore the total mass of the first-level KK-particles according to \eqref{Delta2}:

\begin{eqnarray}\label{densKK}
m_{KK} = \frac{\lambda_1^2 M_P^2 |A|^2 I[\beta] N_{inf}}{96\pi^3 H^4 r_0^3} \exp\left\{3N_{inf}-\frac{2\lambda_1^2N_{inf}}{3H^2r_0^2}\right\},
\end{eqnarray}
where $N_{inf}$ is the total duration of inflation in terms of e-folds. Here we also use the connection: $M_P^2=4\pi r_0^2 m_6^4$ \eqref{denot0}.

The first KK-excitation is protected from decay by conservation of the internal charge. Because of this, it is considered as a particle of dark matter in many models \cite{2003NuPhB.650..391S,cheng2002kaluza}. In our case, the generation of such particles occurs during inflation by breaking the internal symmetry, which prevents them from decay.

\begin{figure}[t]
 \begin{center}
  \subfigure{     \includegraphics[width=0.47\textwidth]{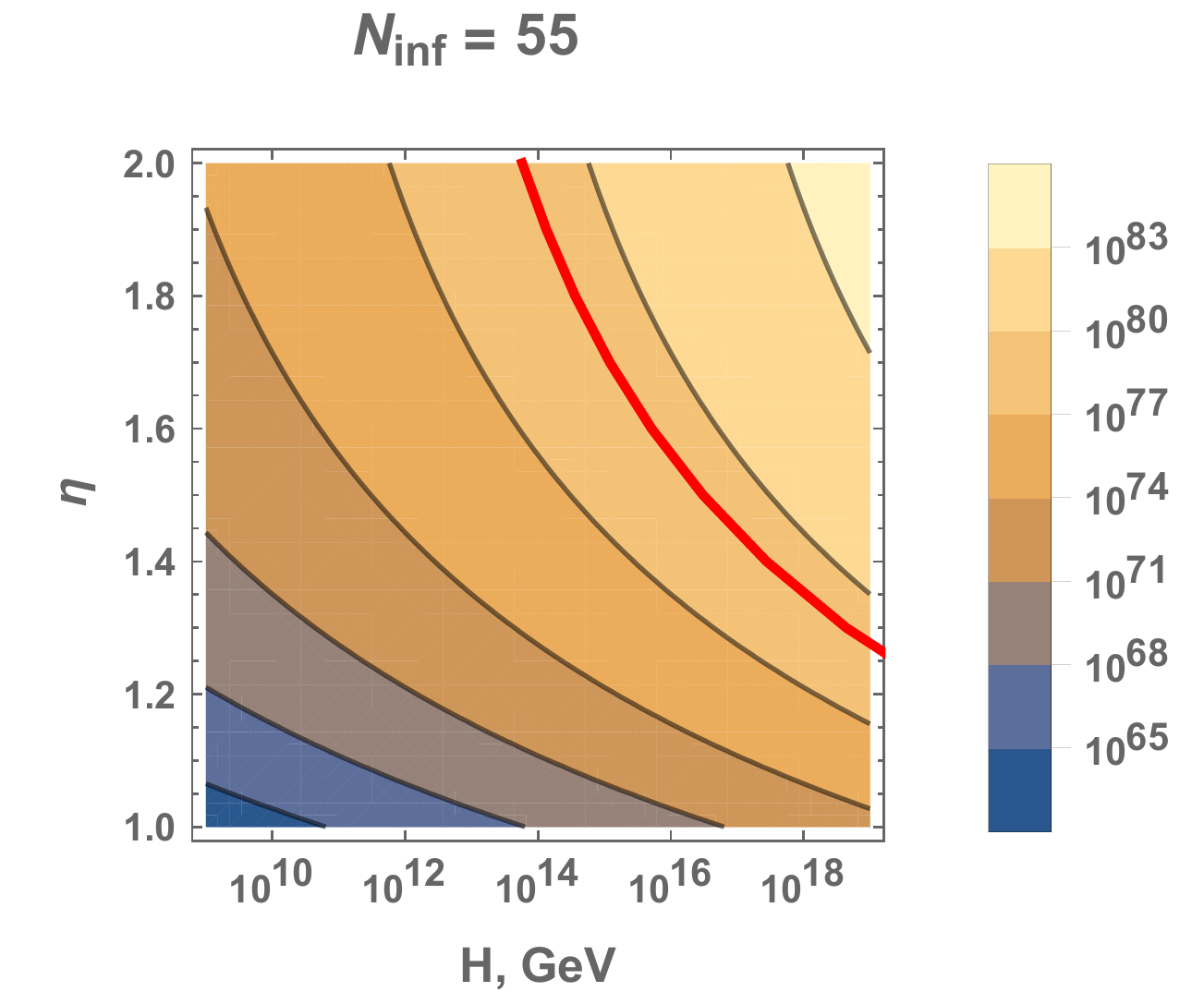}
   \label{fig:55}
  }
  \subfigure{    
   \includegraphics[width=0.47\textwidth]{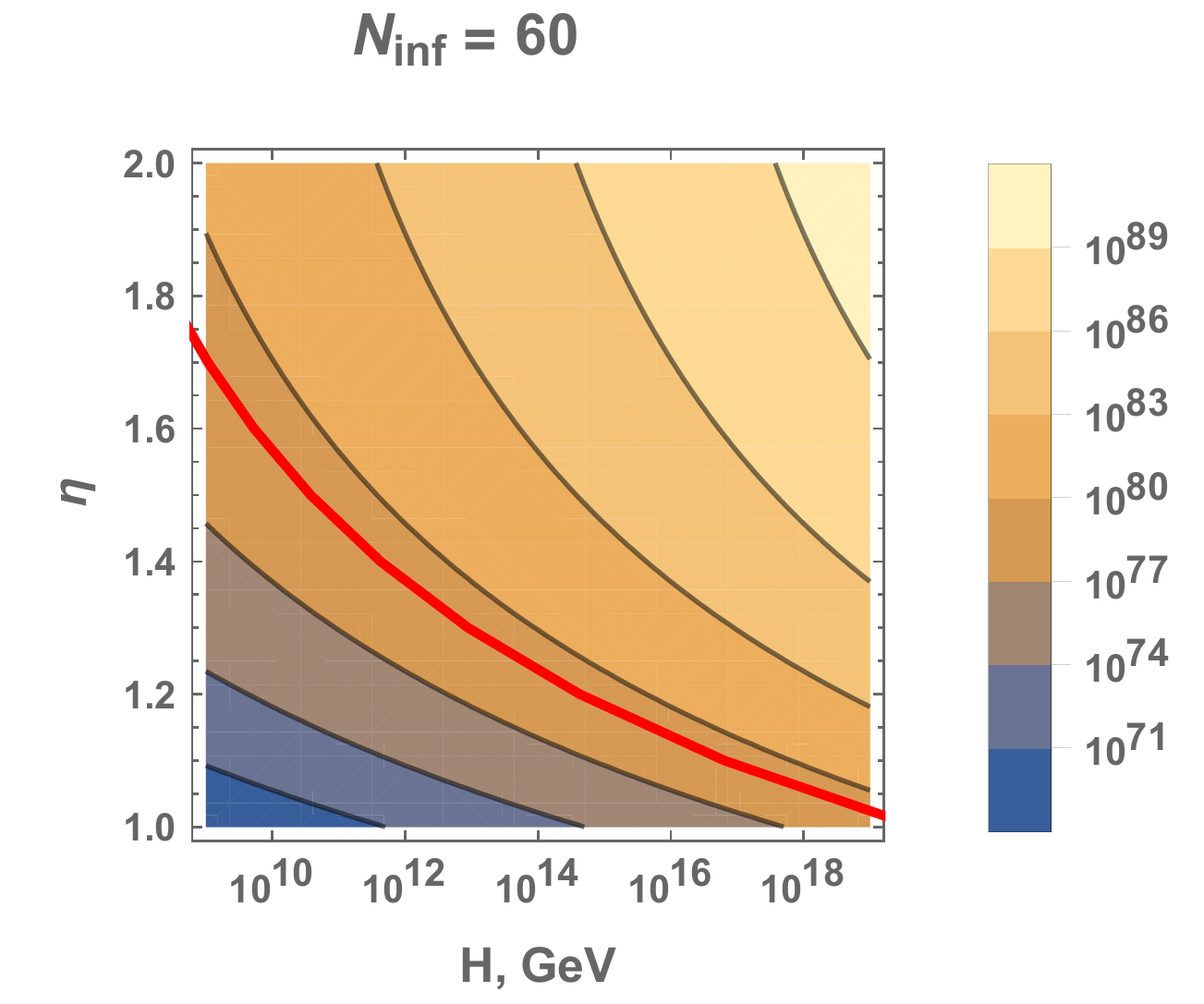}
   \label{fig:60}
  }
 \caption{The total mass (in GeV) of KK-particles in the Universe  depending on $H$ and $\eta=Hr_0$. For comparison, the red line shows the total mass of dark matter in observable Universe ($\sim10^{79}$ GeV) for $N_{inf}=55$ and $60$ e-folds.}
 \label{fig:masskk}
 \end{center}
\end{figure}

Let us estimate the total mass of KK particles assuming that $\lambda_1^2=1$, $I[\beta]=1/\pi$ (similar values were also obtained by numerical simulation of $I[\beta]$). During inflation, the characteristic scale of the dimensionless field amplitude $A \sim H/M_{P}$. The ratio $\eta=H r_0$ strongly affects the numerical value of \eqref{densKK}. As discussed earlier, the case $H \gg 1/r_0$ may lead to contradictions, but $H \gtrsim 1/r_0$ is necessary to excite the first KK-level. We choose $\eta\in[1,2]$ to excite only the first KK-level. The Hubble parameter $H$ may be different depending on the inflationary model, so we vary it in a wide range. According to all the above and expression \eqref{densKK}, the total mass of KK particles is presented in Fig.\ref{fig:masskk}.

As can be seen, the total mass of KK-particles can vary by many orders of magnitude depending on the model parameters. However, a too large total mass may contradict observations. The parameter space above the red line in Fig.\ref{fig:masskk} is forbidden by the observations.

\section{Conclusion}

In this paper, we discuss limits on the extra space size. Modern limits imposed by the collider experiments on the size of extra space dimensions is $l_d< 10^{-18}$cm. The $D$-dimensional Planck mass is limited from below by the value $\sim 10^4$ GeV. As we have shown, the inflationary period gives much more serious boundaries - $10^{-27}$cm and $10^{13}$GeV correspondingly.
The estimation of the compact extra space size remains the same even if this subspace has no any symmetries. 

Particular interest is the case when the size of extra dimensions is of the order of the horizon size, $l_d \sim H^{-1}$. 
In this case, the mass of the first KK-excitation of the order of the Hubble parameter. Hence, any fields like the Higgs field, the inflaton field and so on can be excited during the inflationary state. These excitations are stored during inflation and can be observed nowadays.

\section{Acknowledgement}
The work fulfilled in the framework of MEPhI Academic Excellence Project (contract N 02.a03.21.0005, 27.08.2013)
and according to the Russian Government Program of Competitive Growth of Kazan Federal University. The work of S.G.R. was also supported by the Ministry of Education and Science of the Russian Federation, Project N 3.4970.2017/BY.

\bibliographystyle{unsrturl}
\bibliography{main.bbl}
\end{document}